# Broadband Dual Frequency Comb Spectroscopy in a Rapid Compression Machine

Anthony D. Draper,[1,†] Ryan K. Cole,[1,†] Amanda S. Makowiecki,[1] Jeffrey Mohr,[2] Andrew Zdanawicz,[2]

Anthony Marchese,[2] Nazanin Hoghooghi,[1] and Gregory B. Rieker[1*]

[1]*Precision Laser Diagnostics Laboratory, Department of Mechanical Engineering, University of Colorado Boulder, Boulder, CO 80309, USA*
[2]*Engines and Energy Conversion Laboratory, Department of Mechanical Engineering, Colorado State University, Fort Collins, CO 80523, USA*
*Corresponding author: greg.rieker@colorado.edu
[†]Both authors contributed equally

November 30, 2018

## Abstract

We demonstrate fiber mode-locked dual frequency comb spectroscopy for broadband, high resolution measurements in a rapid compression machine (RCM). We apply an apodization technique to improve the short-term signal-to-noise-ratio (SNR), which enables broadband spectroscopy at combustion-relevant timescales. We measure the absorption on 24345 individual wavelength elements (comb teeth) between 5967 and 6133 $cm^{-1}$ at 704-µs time resolution during a 12-ms compression of a $CH_4$-$N_2$ mixture. We discuss the effect of the apodization technique on the absorption spectra, and apply an identical effect to the spectral model during fitting to recover the mixture temperature. The fitted temperature is compared against an adiabatic model, and found to be in good agreement with expected trends. This work demonstrates the potential of DCS to be used as an in situ diagnostic tool for broadband, high resolution, measurements in engine-like environments.

## 1. Introduction

Infrared laser absorption spectroscopy is a useful technique for quantitative, nonintrusive measurement of gas temperature and species concentration in combustion systems [1]. Absorption spectroscopy measures the amount of light absorbed at specific wavelengths that correspond to rotational-vibrational transitions of the molecules in a gas mixture in order to determine the molecular populations in various quantum states. These populations can be linked back to temperature and species concentrations. The dynamic and harsh conditions found in many high-speed combustion systems place demanding requirements on laser absorption sensors. Specifically, the sensor must be able to quantify the absorption at high sensitivity and on short timescales. The ideal sensor would achieve these goals while also measuring over a broad frequency range at high resolution. Broad spectral bandwidth enables the measurement of multiple species that absorb at different frequencies, increases the temperature sensitivity and range by probing many rotational-vibrational energy levels, and allows the sensor to resolve absorption features over a large range of pressures and through spectral overlap of neighboring absorption transitions. This last capability is crucial as small molecules will exhibit narrow absorption features at ambient conditions which broaden and blend as the pressure increases.

In this paper, we demonstrate sub-millisecond-time-resolved, broadband dual-comb spectroscopy (DCS) with fiber mode-locked frequency combs at the high spectral resolution required to fully resolve absorption features across the full range of temperature and pressure conditions encountered in an engine-like environment. We recover temperature by fitting $CH_4$ absorption features spanning 166 $cm^{-1}$ from 5967-6133 $cm^{-1}$ at 0.0068 $cm^{-1}$ tooth spacing, for a total of 24345 comb teeth. The absorption spectra are obtained at a time resolution of 704 µs in a rapid compression machine (RCM) compression operating from 1 to 21 bar and 294 to 566 K. The power-per-comb-mode is optimized and an apodization technique is used in Fourier processing to improve the short-term SNR. The apodization method introduces an instrument line shape, but this instrument line shape is known exactly, and can be unambiguously accounted for in the spectral fitting routine. This approach demonstrates the bandwidth and resolution benefits of mode-locked dual-comb spectroscopy at the timescales required for rapidly changing combustion environments.

*1.1 Rapid Compression Machines*

RCMs are typically used for laboratory studies of fuels and combustion kinetics, as well as sensor validation in realistic, engine-like conditions [2]. RCMs generally utilize one or more pneumatically driven pistons that, when released, rapidly compress a gas sample with a compression ratio of 10 to 20, resulting in elevated compressed pressures and temperatures (e.g. 10 to 30 bar, 500 to 1000 K) The compression process can occur in as little as 2 ms, and thus is often considered analogous to a single compression stroke of an internal combustion engine [2]. The combination of short timescales with rapidly varying thermodynamic conditions provides a realistic and demanding system in which sensors can be tested and validated.

The temperature evolution during the compression process is of particular interest for chemical kinetics studies due to its relation to reaction rate constants [1]. The temperature during the compression event $T$, can be approximated by Eq. (1) from a known initial temperature $T_0$, initial pressure $P_0$, the measured pressure during the compression $P(t)$, and the ratio of specific heats of the gas mixture $\gamma$ [3].

$$\ln\left(\frac{P(t)}{P_0}\right) = \int_{T_0}^{T(t)} \frac{\gamma}{\gamma - 1} \frac{dT}{T} \qquad (1)$$

This equation relies on the so-called adiabatic core assumption, namely that the compression occurs on such a short timescale that there is negligible heat transfer from the gas. This assumption will lose validity due to aerodynamic mixing during the compression as well as due to any chemical reactions that may occur [3-4]. After a compression the pistons reach their final position, and the gas has additional time for heat transfer with the RCM walls, further deviating from the idealized process. Consequently, accurate temperature measurements are vital for combustion studies both in the time period of the compression process, as well as after the end of compression when the pistons are in their final position.

*1.2 Laser absorption spectroscopy in engine-like environments*

Many laser absorption sensors with narrow wavelength coverage have been demonstrated in rapid compression and shock tube environments with high sensitivity and time resolution [1,5]. Narrowband sensors observe a few select absorption features of a single species that have favorable temperature dependence within a range of interest. As the spectral bandwidth is small, pressure broadening effects can cause the absorption feature linewidth to exceed the measurement range. Together, these features create bounds on the species coverage and usable temperature and pressure ranges of the laser sensor.

Recently, a number of studies have published broadband absorption spectrometers to measure temperature within harsh environments such as an RCM. Because broadband spectroscopy observes many absorption features, these techniques circumvent the limitations imposed by temperature sensitivity of particular features and pressure broadening that challenge narrowband sensors. The broadband nature additionally raises the possibility of multispecies monitoring. These sensors have relied upon external cavity diode lasers [6] or techniques such as Fourier domain mode locking [7] and supercontinuum generation [8,9] to record absorption spectra spanning hundreds of wavenumbers. Notably, Werblinski *et al.* demonstrate supercontinuum absorption spectroscopy in fired [8] and non-fired [9] RCMs. The authors record spectra spanning more than 500 cm$^{-1}$ at a rate of 10 kHz to measure time histories of temperature, pressure, and water vapor mole fraction in the compressed gas through a first derivative fitting approach. So far, broadband approaches in rapidly varying environments have limited spectral resolution (e.g. ~0.8–2.0 cm$^{-1}$ [6–9]) that can limit the published results to pressures above 6 bar, below which the spectrometer resolution matches or exceeds the linewidth of the absorption features.

*1.3 Dual frequency comb spectroscopy at short timescales*

Frequency combs are laser sources that emit many discrete, evenly spaced frequencies of light, often referred to as comb teeth. DCS is an emerging technique that uses two frequency combs and is capable of simultaneous broad bandwidth and high resolution [10]. In DCS, a frequency comb is interfered with a second comb having slightly different tooth spacing in a massively parallel optical heterodyne approach. This allows absorption to be resolved tooth-by-tooth, and for the data to be collected in the RF-domain with a single photodetector and no moving parts.

There are several classes of frequency comb that are currently capable of robust, portable operation: fiber mode-locked, modulator, and quantum cascade laser (QCL) combs. Mode-locked combs exhibit a train of short pulses comprising many thousands of comb teeth with optical frequency spacing matching the pulse repetition rate of the laser [11]. The intensity of the pulses lend themselves to nonlinear processes, and thus mode-locked combs are capable of nonlinear broadening to thousands of wavenumbers. Typically, the comb tooth spacing is very fine (0.003-0.03 cm$^{-1}$), as larger spacing requires high pulse repetition rate, which in turn would lead to short laser cavities and reduced peak power for nonlinear broadening.

Modulator combs are generated by electro-optic modulation of continuous-wave lasers [12,13]. The tooth spacing is given by the frequency of the RF source driving the modulators, thus the spacing can be tuned from $10^{-4}$ cm$^{-1}$ to 1 cm$^{-1}$. However, the bandwidth of these combs is typically limited to a few wavenumbers and requires more complex approaches to achieve a broad bandwidth [14]. QCL frequency combs are semiconductor sources emitting in the mid-IR and THz. They have modest bandwidth (40-100 cm$^{-1}$) in the mid-IR (1111-1667 cm$^{-1}$), with comparably large mode spacing (0.25-0.5 cm$^{-1}$), which leads to a relatively small number of comb teeth across the spectrum (a few hundred).

For a single dual-comb spectrum, the SNR in most cases is limited by the dynamic range of the detection system. In practice, the comb power striking the detector must be reduced to below the point where nonlinearities arise near detector saturation. Thus optimizing the power-per-comb-tooth striking the detector achieves the best noise characteristics. Comb systems operating with a small number of modes over a limited range, such as QCL and most modulator combs, are therefore capable of high short-term SNR at the expense of resolution and/or bandwidth. Mode-locked combs with broad bandwidth and close tooth spacing place many more comb modes on the detector, spreading the power-per-mode and reducing short-term SNR, but with much higher bandwidth and resolution. At long timescales, the achievable SNR for DCS is driven by the ability to coherently average measured spectra. The ultimate SNR is therefore driven more by comb stability or the ability for software to correct for stability. Consequently, absorption spectra for mode-locked DCS are typically averaged on the order of seconds to minutes to reach very high SNR. These constraints have generally positioned femtosecond mode-locked comb sources for broadband multispecies measurements in applications with relaxed time resolution.

A representative diagram of mode-locked DCS is shown in Figure 1. DCS with mode-locked lasers is most easily described in the frequency domain, where each comb spectrum consists of evenly spaced frequency modes (teeth). Two frequency combs are phase locked to a stable reference such that the signal comb has repetition rate $f_{rep,S}$ and the second, local oscillator, comb has repetition rate $f_{rep,LO}$, where $f_{rep,S} - f_{rep,LO} = \Delta f_{rep}$. Moving across the spectrum, the optical frequency of each pair of fast and slow comb teeth differ by a frequency $n\Delta f_{rep}$, where $n$ is an integer number of tooth pairs from the point where the frequencies of a pair of teeth exactly overlap. This difference frequency is detected on a photodiode as a heterodyne beat frequency, effectively mapping the magnitude of each pair of comb teeth from the optical domain into the RF-domain as shown in Figure 1d.

DCS can be described equivalently in the time domain as two femtosecond pulse trains with different repetition rates. In the frame of a pulse emitted by the signal comb, pulses from the local oscillator comb iteratively step through the signal comb pulse at an effective time step of $\Delta T_E = \Delta f_{rep}/(f_{rep,S} f_{rep,LO})$ [15]. These pulse trains are incident on a photodetector, such that they form a cross-correlation between the electric fields of the different comb pulses (termed an interferogram), that can be seen in Figure 1b, which exhibits an intense centerburst when the pulses of the combs are overlapping in time. The interferogram is digitized by commercial electronics with a laboratory time step between data points of $\Delta T_L = 1/f_{rep,LO}$. Each interferogram is constructed of $N = f_{rep,LO}/\Delta f_{rep}$ individual points, and thus the laboratory time to acquire a single interferogram is given by $t_{acq} = 1/\Delta f_{rep}$. The frequency domain spectrum described in the preceding paragraph is the Fourier transform of the time domain interferogram. A more detailed description of the DCS method applied here is given in [10,15,16].

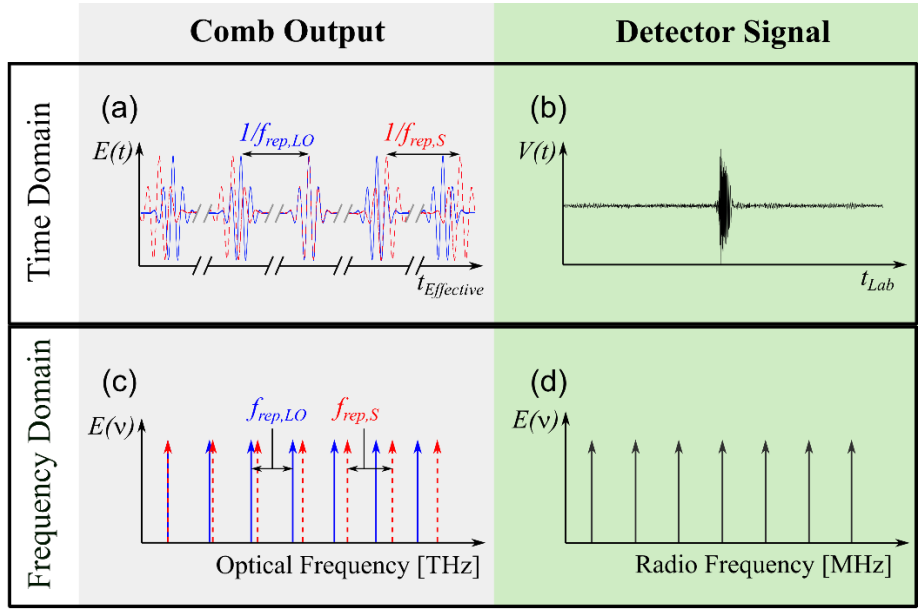

Fig. 1. Diagram showing the time and frequency domain representation of dual frequency comb spectroscopy. (a) Effective time domain schematic of femtosecond comb pulses originating from combs with different repetition rates, where the pulses of one comb "step through" the pulses of the other comb. The signal comb is shown in dashed red and the local oscillator comb in blue. (b) The time domain signal as recorded by the photodetector is a cross-correlation of the comb pulses – an interferogram – formed in laboratory time and with a "centerburst" when the pulses of the two combs are overlapping. (c) The comb teeth of the two combs represented in the optical frequency domain. The two comb repetition rates are fixed such that a unique heterodyne beat in the RF-domain is formed from pairs of comb teeth. (d) Individual comb teeth are resolved in the RF-domain by taking the Fourier Transform of the interferogram shown in (b).

As mentioned above, the short term SNR of this detection method is typically limited due to the low optical power-per-comb-mode. As such, previous practical deployments of mobile, mode-locked dual-comb spectrometers [17,18] typically average for 60-120 seconds to build to high SNR. In order to resolve transient combustion phenomena, this averaging time must be decreased by upwards of a factor of 10,000. The first step to increasing the short-term SNR is to optimize the incident power on the photodetector. The approach is to maximize the power-per-comb-tooth (and thus the SNR) on the detector by filtering all comb teeth outside of the spectral region of interest, and to optimize the optical power incident on the photodetector such that it is operated just below the power level where detector non-linearity becomes large.

The short-term SNR for mode-locked DCS can be further increased by applying an apodization technique to the measured interferograms in post-processing. Most of the signal of interest in DCS is collected during the centerburst of the interferogram: when the pulses from the two combs are overlapping in time on the detector. The rest of the interferogram is still important to collect in order to obtain the right number of points for a full, comb-tooth-resolved spectrum, and because one must wait for the pulses of the two combs to overlap again. Apodization multiplies a function with the interferogram in order to retain the signal around the centerburst while filtering out the noise on the rest of the interferogram, which would subsequently become noise in the frequency domain. However, the cost of this technique is the addition of an instrumental line shape in the resulting absorption spectrum [19]. In this work, we apodize the interferograms with a boxcar function to increase the short-term SNR. The instrument line shape resulting from a boxcar apodization function is described analytically as Eq. (2) where $\Delta_E$ is the HWHM of the boxcar function in effective time, and $v$ is the frequency.

$$f(v) = 2\Delta_E \text{sinc}(2\pi\Delta_E v) \qquad (2)$$

Thus, the instrument line shape is perfectly known, and can therefore be applied to the absorption model that is used to fit the data. In order to demonstrate the feasibility of this method to a fitting routine with a spectral database, a 1 bar and 300 K CO line near 6297.5 cm$^{-1}$ was measured in a static configuration with the fiber mode-locked dual-comb spectrometer discussed in the following section. A boxcar apodization was applied to both the experimental data and the HITRAN 2016-based absorption model. Figure 2 shows the agreement between the apodized model and measurement. As the instrument line shape is a known analytical function, the model replicates the effects of apodization exactly. Therefore, the outcome of a fitting routine used to determine the temperature or species concentration is unbiased by the addition of the instrument line shape.

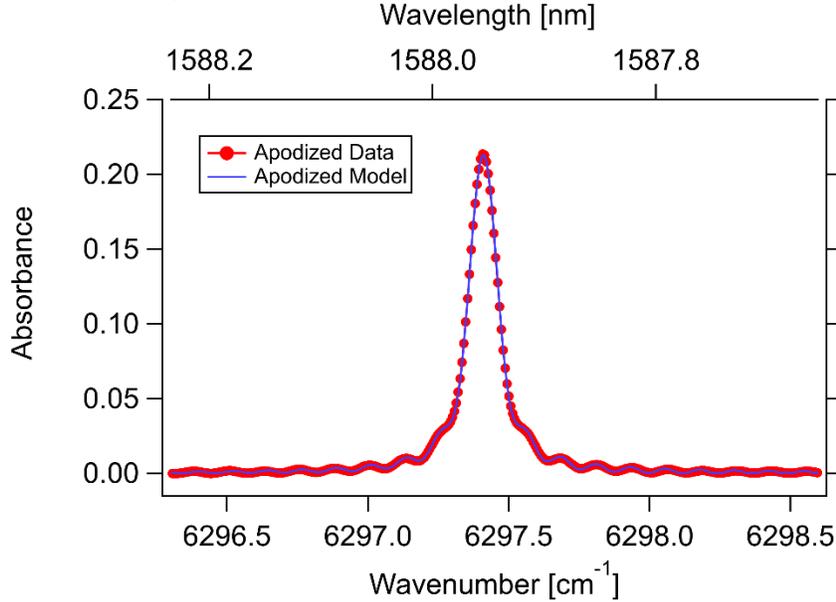

Fig. 2. Measured CO absorption feature near 6297.5 cm$^{-1}$, with the same apodization applied to the measured interferogram (red) and the model (blue). The instrument lineshape from apodization is apparent, but is well captured by the model.

2. **Experimental setup**

The RCM employed in this work was manufactured by Marine Technologies Ltd., and is housed in the Engines and Energy Conversion Laboratory at Colorado State University (CSU). The RCM employs dual piston design with creviced pistons that reduce the effects of aerodynamic mixing during the compression process [3,20], and was configured for a compression ratio of 12.5. Optical access to the combustion chamber is provided through two sapphire windows with a 3° wedge to discourage etalon interference effects. Gas pressure is measured before the compression process using an Omega DPG409 pressure transducer (accuracy 0.08% of reading), and monitored during the compression stroke using a Kistler 603B pressure sensor (accuracy 1% of reading). A detailed description of the CSU RCM facility can be found in [20,21].

A near-IR mobile dual-comb spectrometer [17,18,22] was phase locked with $\Delta f_{rep} = 2837\ Hz$, giving a single spectrum acquisition time of 352 µs. The laser light was filtered to cover 5967-6133 cm$^{-1}$, and an amplifier boosted the power in this frequency region up to the optimal range for the detector. The point spacing of the spectrum is set by the laser repetition rates to be 0.0068 cm$^{-1}$, which results in 24345 individually resolved comb teeth accessible for fitting. Light from the DCS was delivered with singlemode fiber to the RCM, collimated, and passed though the 4.63 cm combustion chamber. Laser light leaving the RCM is incident on a DC-coupled fast photodetector (ThorLabs PDA10CF), low-pass filtered (100 MHz bandwidth), and recorded with a 250 MS/s digitizer. A schematic of this experimental configuration is shown in Figure 3.

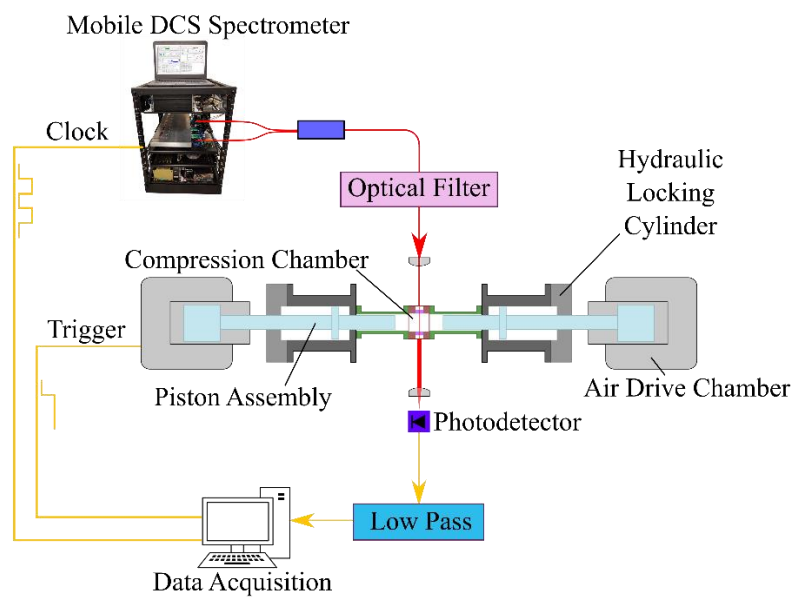

Fig. 3. Schematic of experimental setup. The mobile DCS supplies light via single-mode fiber to a collimator, which transmits the light through the RCM combustion chamber, and is then received through a convex focusing lens onto a fast photodetector. The resulting signal is low-pass filtered before being collected with the data acquisition system. The spectrometer also provides a clock signal to the data acquisition, and the RCM start switch is linked to enable synchronous collection.

The RCM combustion chamber was maintained at 294 ± 0.5 K, the pistons retracted and locked, and the chamber evacuated to 6.9 ± 0.8 mbar. A vacuum dataset was collected to aid in baseline correction in post-processing before compression. Subsequently, $CH_4$ was first added to the chamber, and then $N_2$, to give an initial pressure of 1001 mbar, with a $CH_4$ mole fraction of 0.750. Data collection was triggered synchronous to the RCM pistons being released via a common start signal. Time-domain interferograms were collected during the compression process through an automated LabVIEW program.

## 3. Results and analysis

### 3.1 Data processing

Every two interferograms measured through the RCM were averaged for an overall measurement time resolution of 704 µs. A boxcar apodization HWHM of $\Delta_E = 229\ ps$ was applied to each binned interferogram, yielding a FWHM of the instrument line shape of 0.088 cm$^{-1}$. The spectral resolution with the instrument line shape can be estimated from Rayleigh's definition as $1/\Delta_E = 0.15\ cm^{-1}$ [15,19], which yields the experimental spectral resolution. This degree of apodization was selected in order to maintain the ability to resolve the $CH_4$ absorption features at 1 bar and 298 K before the RCM compression process.

Apodized interferograms were subsequently Fourier transformed to yield the $CH_4$ transmission spectrum through the RCM. Each transmission spectrum was baseline corrected using the vacuum background dataset and converted to absorbance units with Beer's law. A polynomial baseline fitting routine (described in [23] and the supplement to [24]) was used to correct residual baseline irregularities caused by subtle differences between the vacuum background and compression datasets. Experimental spectra at the initial conditions are shown in Figure 4 to illustrate the effect of apodization. The resulting SNR benefit is dramatically evident. From the zoom view shown in Figure 4b, one can see the emergence of the $CH_4$ absorption feature from the noise using the apodization approach.

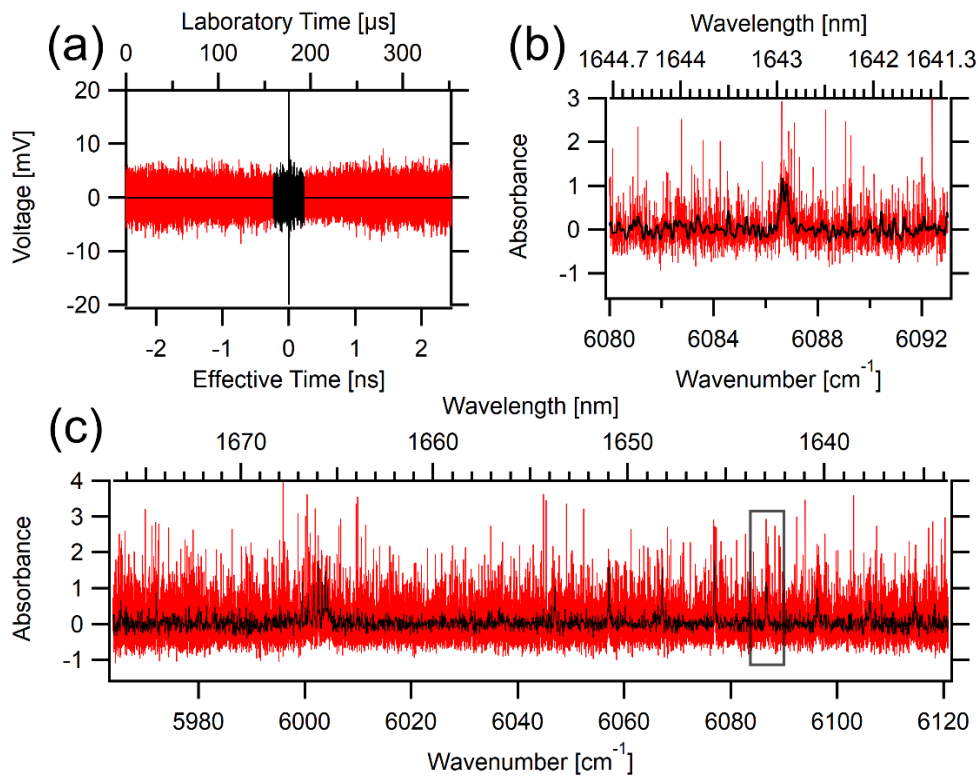

Fig. 4. Measured absorbance spectra at 1 bar and 294 K with 704 μs time resolution just prior to the compression. The red traces are the unapodized data and the black traces are apodized. (a) Interferogram with both the laboratory and effective timescales shown, (b) Zoomed $CH_4$ absorption feature for the region indicated on panel c, and (c) Full measured spectrum.

*3.2 Temperature Fitting*

A Levenberg-Marquadt algorithm was utilized to fit a model generated from the HITRAN 2016 database [25] with the Python HAPI package [26] to each baseline corrected spectrum. The mixture mole fraction and pressure were held constant and temperature was allowed to vary (with an initial guess from GASEQ calculations, described later). The resulting temperature fit is shown in Figure 5.

Using the measured pressure, the compression temperature was calculated with GASEQ, a thermodynamic equilibrium software [27]. The measured temperature tracks well with the calculation during the compression process, 0 to 15 ms, and reaches a peak temperature 6 K lower than the 566 K predicted by GASEQ. This agreement is likely within the uncertainty of the GASEQ calculation, which we do not directly estimate here, but is affected by uncertainty in the initial temperature, gas composition, and measured pressure. The dual-comb measured temperature decreases faster than the calculation once the pistons lock at the end of compression, ultimately reading approximately 19 K lower than the GASEQ predicted temperature after 25 ms. This deviation could arise from several effects. For example, heat loss and gas slippage from the compression chamber are not accounted for in GASEQ, though they do affect the result through the measured pressure. In addition, the formation of cold boundary layers can influence the path-averaged measured temperature to a value that is slightly different than the true average. With further detailed consideration of the results, it should be possible to determine the rate of heat loss. Having a measure of the rate of heat loss from the system also allows for system calibrations and compensations in subsequent RCM experiments.

The systematic uncertainty of the frequency comb measurement is represented by the uncertainty bars on Figure 5. The uncertainties of the pressure transducers, the physical tolerance stack-up of the path length, and the initial temperature were included as input values into the fitting routine to determine their influence on the measured temperature values. At the initial onset of the compression, the uncertainty is ±6 K, which increases to ±20 K at the end of data collection. The increase is a result of increasing pressure transducer uncertainty and the greater influence of the uncertainty in the initial conditions as one scales to higher compression.

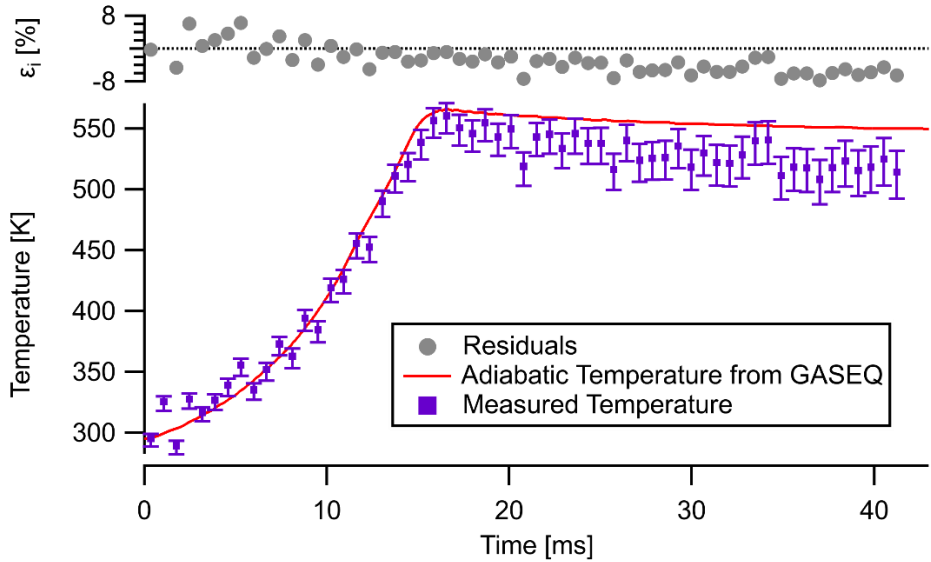

Fig. 5. DCS-measured temperature from broadband fitting of the apodized spectra (purple). The red trace is the adiabatic temperature calculated using GASEQ, while the gray represents the residual percent error.

The precision of the measurement was assessed using the residual between the measured and calculated values. The residual percent error was split into two sets: values leading up to and including the peak compression, and values from the peak until the end of the data set. The linear trends were removed from the residuals, the normality of the two sets verified through the Shapiro-Wilk test, and their standard deviations calculated. Up until the compression peak, the standard deviation of the residual percent error is ±3.1%, and ±1.5% afterwards. In the relatively constant temperatures following the compression peak, the 1.5% residual percent error corresponds to approximately ±8.5 K measurement precision.

The high precision and low uncertainty is due in part to the broadband nature of the comb – the fit includes a very large number of comb teeth probing a large number of $CH_4$ absorption features, both of which lead to a robust fit. The dual-comb spectrometer is able to precisely resolve the $CH_4$ absorption features from atmospheric pressure all the way to the peak pressure of 21.15 bar.

## 4. Conclusions

We demonstrate high-speed, broadband, mode-locked dual frequency comb spectroscopy in an RCM by measuring the temperature of a $CH_4$-$N_2$ gas mixture with 704 μs time resolution. Absorption features were measured with 24345 comb teeth between 5197-6133 $cm^{-1}$. The short-term SNR of the portable spectrometer was improved through optical power optimization and the application of a boxcar apodization function to the interferograms in post-processing. The apodization technique introduced an exactly known instrument line shape, with a theoretical spectral resolution of 0.15 $cm^{-1}$. With this resolution, absorption features are fully resolved throughout the full range of temperatures and pressures encountered in the compression. The compressed gas temperature was measured using a broadband fitting algorithm with a model that incorporated the exact instrument line shape introduced from apodization. The fitted temperatures agree well with the predicted adiabatic values calculated from the measured pressure. This comparison shows a 6 K under prediction of the 566 K peak as well as indications of the breakdown of the adiabatic assumption after the end of compression. These results indicate that mode-locked DCS can serve as a powerful diagnostic tool for broadband, high resolution spectroscopy in transient combustion environments.

The techniques presented in this paper will see enhanced applicability when combined with rapidly evolving mid-IR mode-locked dual-comb technology. These spectrometers will leverage the bandwidth and fine mode spacing of existing DCS systems with higher sensitivity afforded by the stronger absorption cross sections characteristic of the mid-infrared. The increase in sensitivity will enable temperature measurements at higher time resolution, while the high bandwidth will enable multiple combustion-relevant species to be measured with a single spectrometer.

**Funding.**

Air Force Office of Scientific Research Grant (FA9550-17-1-0224), NASA Fellowship (18-PLANET18R-0018), Defense Advanced Projects Agency (W31P4Q-15-1-0011), and the National Science Foundation (CBET 1454496). The views and conclusions contained in this document are those of the authors and should not be interpreted as representing the official policies, either expressed or implied, of the Defense Advanced Research Projects Agency, the U.S. Army, or the U.S. Government.

**Acknowledgments**

We would like to thank Dr. Colin Gould and Dr. Azer Yalin for their help preparing for the RCM experiment.